\documentclass[%
 reprint,
nofootinbib,
 amsmath,amssymb,
prc, 
]{revtex4-1}

\usepackage{graphicx}
\usepackage{dcolumn}
\usepackage{bm}
\usepackage{caption}
\usepackage{subcaption}
\usepackage[usenames, dvipsnames]{color}

\begin{document}

\preprint{APS/123-QED}

\title{Modeling magnetic neutron stars: a short overview}

\author{R.O. Gomes}
\affiliation{Frankfurt Institute for Advanced Studies,
Frankfurt am Main, Germany}
\email{rosana.gomes@ufrgs.br}
\author{V. Dexheimer}
\affiliation{Department of Physics, Kent State University, Kent OH 44242 USA}
\author{S. Schramm}
\affiliation{Frankfurt Institute for Advanced Studies,
Frankfurt am Main, Germany}

\date{\today}

\begin{abstract}
Neutron stars are the endpoint of the life of intermediate mass stars and posses in their cores matter in the most extreme conditions in the universe. Besides their extremes of temperature (found in proto-neutron stars) and densities, typical neutron star' magnetic fields can easily reach trillions of times higher the one of the Sun. Among these stars, about $10\%$ are denominated \emph{magnetars} which possess even stronger surface magnetic fields of up to $10^{15}-10^{16}\,\mathrm{G}$.
In this conference proceeding, we present a short review of the history and current literature regarding the modeling of magnetic neutron stars. Our goal is to present the results regarding the introduction of magnetic fields in the equation of state of matter using Relativistic Mean Field (RMF) models and in the solution of Einstein's equations coupled to the Maxwell's equations in order to generate a consistent calculation of magnetic stars structure. We discuss how equation of state modeling affects mass, radius, deformation, composition and magnetic field distribution in stars and also what are some of the open questions in this field of research.   \end{abstract}

\pacs{Valid PACS appear here}
\maketitle


\section{\label{Intro}Introduction}

In 1979, when the \emph{gamma-ray burst} detection was already ongoing, a peculiar explosion occurred. It was characterized by a short burst of high energies (\emph{hard $\gamma$-rays}) followed by a soft $\gamma$-ray emission that decayed with a sinusoidal variation. The explosion was so extreme that for a fraction of seconds saturated all $\gamma$-ray detectors in orbit. In the following decade, similar $\gamma$-ray explosions were detected and the sources identified as small remnants of supernovae \cite{Duncan:1992hi}, making it possible conclude that a new class of objects had been found, named \emph{soft gamma repeaters (SGRs)}, though the mechanism responsible for generating such explosions remained unknown.

With the advances on $X$-ray detection, new objects with peculiar radiation emission were also observed and, in particular, another class of objects denominated \emph{Anomalous X-ray Pulsars (AXPs)} was identified. AXPs present low rotation rates ($P\sim 6-12\,\mathrm{s}$) and high surface magnetic fields ($B\sim 10^{13}-10^{15}\,\mathrm{G}$). They are called \emph{anomalous} because their X-ray radiation could not be explained by  their rotation rates, as typical pulsars are.

The further monitoring of SGRs and AXPs made it possible to characterize that both these objects present low rotation rates and high magnetic fields \cite{Kouveliotou:1998ze}, though their radiation emission behavior were distinct.
SGRs present gamma-ray explosions with repetition periods that can vary from seconds to years. 
During the inactive phase, a steady X-ray emission 
is also identified.
Furthermore, SGRs also produce uncommon 
and extremely energetic explosions denominated \emph{giant flares}, which can be up to ten times more luminous than a supernova event \cite{ferro_6}.
AXPs are characterized by their steady X-ray emission, but in the scale of hard X-rays. 
Besides the X-ray emission, also present \emph{glitch} phenomena, in which an increase in the spinning rate, followed by a return to the original spin rate is observed \cite{Livingstone:2007bn}. AXPs differ from SGRs due to their lack of activities such as periodic emissions or giant flares explosions. 

The current interpretation is that these objects are two different life stages of objects denominated \emph{magnetars} \cite{Kaspi:2016jkv,duncan_website}, which are magnetically powered neutron stars. The mechanism responsible for their radiation emission comes from the strong magnetic field decay along time \cite{Thompson:1995gw,Thompson:1996pe}, being SGRs associated to young magnetars and AXPs to older ones. SGRs present high energy explosions due to instabilities from the magnetic tension between core and crust, or even the large scale rearrangement of magnetic fields (giant flares); while AXPs present steady X-ray emission due to the collision of high energy particles on their surfaces and magnetic field lines (also in SGRs). 

The conservation of magnetic flux during the supernova collapse is not able to explain the intensities of magnetic fields found in magnetars, as the radius of a $1.4\,M_{\odot}$ star would have to be smaller than its Schwarzschild radius in order to generate a magnetic field of $10^{15}\mathrm{G}$ \cite{Tatsumi:1999ab}. 
In 1992, Robert Duncan \& Christopher Thompson proposed a mechanism, denominated \emph{magnetohydrodynamic dynamo mechanism (MDM)}, based on the amplification of magnetic fields through the combination of rotation and convection in hot proto-neutron stars \cite{Duncan:1992hi,Thompson:1993hn}, which is currently the most accepted to explain magnetars. According to this theory, after the proto-neutron star phase, convection stops and stars are left with a magnified magnetic field that can reach up values observed in magnetars, marking the end of the dynamo process.

Currently there are 29 magnetars identified 
\cite{Kaspi:2016jkv} 
and, in particular, recent observation of the source 4U 0142+61 suggests internal magnetic fields of $10^{16}\,\mathrm{G}$ \cite{Makishima:2014dua}. Moreover, the Virial theorem estimates that central magnetic fields can reach up to $10^{18}-10^{20}\,\mathrm{G}$ \cite{Lai1991,Cardall:2000bs,Ferrer:2010wz}.

Even though most high energy phenomena involving magnetars have to do with the modeling of their crust dynamics, understanding the effects of strong magnetic fields in the inner core of these objects is of crucial importance for the big picture of matter's behavior in neutron stars. 
To fulfill this a goal, a self-consistent formalism that describes both microscopic and macroscopic features of these objects is needed. Modeling magnetic neutron stars in the presence of strong magnetic fields involves the modeling of both equation of state (EoS) and solving the coupled system of Einstein-Maxwell's equations. In what follows, we present a brief summary of the current status of this field, focusing on these two topics (EoS and structure). For a more detailed discussion about magnetic field effects in the cooling of neutron stars, see Ref. \cite{Negreiros:2018cjk} and references therein.
\section{\label{EOS} Magnetic fields in the Stellar EoS}
\subsection{\label{landau} Landau Quantization}
Effects of strong magnetic fields in a low-density Fermi gas were first performed by Canuto  \cite{Chiu:1968zz,Canuto:1969ct,Canuto:1969cs,Canuto:1969cn}, followed by calculations using RMF models with magnetic field effects  \cite{Chakrabarty:1997ef,Broderick:2000pe}. 
The introduction of magnetic effects in RMF models is given by adding the electromagnetic interaction term in the lagrangian density of models. This is done under the consideration of a constant dipolar (usually in $z$ direction) and external magnetic field that generates a quantization of charged particles energy levels:
$E_{\nu} =\sqrt{ m^2 + k_z^2 + 2 |q| B \nu }$,
where $m$ and $k_z$ are the mass and Fermi momentum in the $z$ direction of the charged particles, respectively, and $\nu \equiv  l + \frac{1}{2} -  \frac{s}{2} \frac{q}{|q|}$ (for 
$\nu \geq 0$).

The quantum number $\nu$ is associated to the so called Landau quantization of energy levels of charged fermions and, according to the expression above depends on the charge $q$, quantum orbital number $l$ and spin $s$ of particles. One can easily check that besides for the ground state ($l=0$, $s=+1$), all other Landau energy levels are double degenerate. 
Furthermore, under conditions of zero temperature (relevant of neutron stars), energy levels obey Fermi-Dirac statistics, 
having a Fermi energy $E_F$ and a maximum Landau level $\nu_{max} < (E_F^2-m^2)/2|q|B$, beyond which $k_z$ becomes negative. 
Landau quantization introduces a sum over levels $\nu$ for EoS calculations,  running only through the lower Landau levels for strong magnetic fields. In the weak field regime, the sum reaches the continuous limit, making the integrals equivalent to the $B=0$ case \cite{Strickland:2012vu,Negreiros:2018cjk}. 

When magnetic fields are introduced in RMF models, the effect is the softening of the EoS due to the increase of charged particles population and, hence, decrease of isospin asymmetry  \cite{Lai1991,Chakrabarty:1997ef,Broderick:2000pe}.
In addition, magnetic field effects give rise to an anisotropy in the matter energy-momentum tensor components, generating two pressure components \cite{Canuto:1969ct,Canuto:1969cs,Broderick:2001qw,PerezMartinez:2007kw,Strickland:2012vu,Dexheimer:2012mk}: $P_{\parallel} = - \Omega, \quad
P_{\perp} = -\Omega - B {\cal M},$
where $P_{\parallel}$ and $P_{\perp}$ are the pressure components in the parallel and perpendicular directions to the external magnetic field, $\Omega$ is the grand potential and ${\cal M}$ is the magnetization of matter.

Effects of anomalous magnetic moment were also investigated  through the corresponding coupling of particles (charged and uncharged) to the electromagnetic field tensor  \cite{Canuto:1969cn,Broderick:2000pe,Broderick:2001qw,Strickland:2012vu}.
For the case of hadronic matter, it has been concluded that strong magnetic fields turn the EoS stiffer due to polarization effects. However, a significant impact is only identified for fields up to $B\sim 10^{18}\,\mathrm{G}$ \cite{Dexheimer:2012qk}, whereas for quark matter much smaller effects have been estimated so far  \cite{Ferrer:2015wca}. The modeling of hybrid stars under strong magnetic fields is not an easy task, as the matching of two phases has to take into account magnetization. 

In the past, magnetic effects were introduced in many RMF models for describing neutron stars with different population content \cite{Chakrabarty:1997ef,Broderick:2000pe,Broderick:2001qw,PerezMartinez:2007kw,Rabhi:2009ih,Paulucci:2010uj,Sinha:2010fm,Orsaria:2010xx,Dexheimer:2011pz,Dexheimer:2012mk,Dexheimer:2012qk,Lopes:2012nf,Casali:2013jka,Denke:2013gha,Gomes:2014dka,Ferrer:2015wca,Schramm:2015lga}, but still solved the Tolman-Oppenheimer-Volkoff (TOV) equations for  spherically symmetric stars. In such cases, either only the isotropic component of the pressures or the perpendicular component were used as the only pressure contribution for calculating the structure of stars, under the assumption that anisotropy effects are small. However, as we discuss in the next section, deformation effects are extremely important for the  description of neutron stars that present strong magnetic fields. 
\subsection{\label{pop} Particle Population}
As densities increase in the interior of neutron stars, it is expected that new degrees of freedom such as hyperons, delta resonances or even  quark matter can take place in the core of these objects and are, therefore, also affected by strong magnetic fields.
As already discussed in the previous session, magnetic fields affect the energy levels of charged particles and, when AMM is considered, even of uncharged ones. 
The effect of magnetic fields on the particles population of neutron stars is pushing the exotic particles threshold of appearance to higher densities \cite{Gomes:2013sra,Gomes:2014dka}.
However, such effects are shown to be significant only for very strong magnetic fields of the order $B\sim10^{18}\,\mathrm{G}$. 

A more relevant source of exotic particles suppression comes from the substantial reduction on the central density of strongly magnetized stars. As magnetic fields have the effect of opposing gravity, similar to rotation effects, the
central density of stars drops and, depending on the stiffness of the EoS, exotic particles such as hyperons might completely vanish \cite{Gomes:2017zkc}. Similar effects can be identified in the case of hybrid stars \cite{Franzon:2015sya,Franzon:2016urz}.


Because of the substantial change in particles 
population due to magnetic fields, it is predicted a particles re-population (appearance of exotic degrees of freedom) as the magnetic field strength of neutron stars decays through time. Such a transition could become detectable via gravitational wave emission in the future, with  new generations of gravitational wave detectors. 
\section{\label{structure}Magnetic fields in the Stellar Structure}
\subsection{\label{LORENE} Einstein-Maxwell Equations}
Introducing strong magnetic field effects in the  stellar structure of compact stars involves solving the coupled  Einstein-Maxwell equations with a metric that allows stars to be deformed.
For solving such a complicated system of equations numerically, a formalism that uses the 3+1 decomposition of space-time technique was developed by Bonazzola et al. \cite{Bonazzola:1993zz}, and implemented in the so called LORENE (Langage Objet pour la RElativite NumeriquE) library.

Due to the non-negligible pressure anisotropy in the presence of strong magnetic fields, an axisymmetric formalism is necessary for describing the system, meaning that the metric potentials will depend both on the radius and on the angle with respect to the symmetry axis ($r,\theta$).
In the formalism described above, a poloidal magnetic field is assumed, and generated by a current function, which is of the library inputs. It is important to stress that although the current function (which ultimately generates the magnetic field distribution) depends on the EoS for each layer of the star, this formalism does not consider hydrodynamical effects   inside stars. This means that the current function is a free parameter that, together with the EoS, determines the dipole moment of the star, as well as their macroscopic properties. 

Moreover, the energy-momentum tensor is decomposed into two parts: perfect fluid (PF) and purely magnetic field (PM) contributions. When the PM contribution exceed the PF contribution along the symmetry axis of the star, the code stops converging, 
imposing a limit of $B_c\sim10^{18}\,\mathrm{G}$ for the central magnetic fields of stars which does not depend strongly on the EoS model  \cite{Bonazzola:1993zz,Bocquet:1995je}.

This formalism initially was applied to describe magnetic neutron stars without taking magnetic effects into account in the EoS  \cite{Bocquet:1995je,Cardall:2000bs}. 
It was only recently that works considering self-consistent calculations that includes magnetic fields both in the EoS and structure of stars were implemented to describe quark stars, hybrid stars and hadronic stars \cite{Chatterjee:2014qsa,Franzon:2015sya,Dexheimer:2016yqu,Gomes:2017zkc}.
\subsection{\label{macroscopic} Macroscopic Properties}
When a self-consistent approach was used for the first time for describing magnetic neutron stars, it was identified that introducing magnetic fields on the EoS do not impact significantly the macroscopic properties of neutrons stars, such as stellar masses or radii  \cite{Chatterjee:2014qsa,Franzon:2015sya,Gomes:2017zkc}.
However, even though small, an increase in the mass of stars is identified when anomalous magnetic moment effects are introduced in the EoS \cite{Franzon:2015sya}. This comes from the fact that the limit in which magnetic fields become relevant for the EoS for the core of neutron stars is the same as the one for the  convergence limit for the numerical calculations.

Still, strong magnetic fields impact directly the masses (baryon and gravitational) and deformation neutron stars. 
These effects come mainly form the pure magnetic field contribution to the energy momentum tensor, making the stars more massive due to the extra electromagnetic energy available to prevent the stars collapse.
For a fixed stellar baryon mass, the mass increase in magnetic stars is of the order of percents, and depends on the EoS and current function \cite{Franzon:2015sya}.

Another important aspect to be taken into account is the deformation of stars into an oblate shape, which is directly associated to the poloidal magnetic field distribution assumed  \cite{Bonazzola:1993zz,Bocquet:1995je,Franzon:2015sya}. 
The amount of deformation a star gets depends on the choice of current function, but also on EoS. Softer EoS's allow for higher central densities and, consequently, higher central magnetic fields. However, although stars modeled with a softer EoS present higher internal magnetic fields, they are also more compacts, meaning that the larger radii produced by stiffer EoS's generate stars that are more easily deformed \cite{Gomes:2017zkc}. 

It was estimated that neglecting deformation of magnetic stars by solving  TOV equations when determining the structure of stars leads to a $12\%$ overestimation of the maximum mass and a $20\%$ underestimation of equatorial radius of a $1.4M_{\odot}$ star \cite{Gomes:2017zkc}. These results were obtained for the maximum magnetic field limit of the LORENE code. Further detailed calculations for the limit in which a spherically symmetric geometry is valid to model magnetic stars is still needed.
\subsection{\label{mag-distribution} Magnetic Field Distribution}
As already discussed, when a poloidal magnetic field configuration is assumed, the magnetic distribution as a function of density (or chemical potential) depends on the EoS and on the choice of current. In past, when such calculations were not available, ad hoc formulas for magnetic field profiles inside neutron stars were used for introducing magnetic fields on the equation of states of RMF models. 
Such formulas introduced an exponential dependence of the fields with the baryon density, making it possible stars with surface and central magnetic fields of $10^{15}\,\mathrm{G}$ and $10^{19}\,\mathrm{G}$, respectively, to exist. 

Though those magnetic profiles were the only available way to overcome using constant magnetic fields for EoS calculations, they 
do not fulfill Maxwell’s equations \cite{Menezes:2016wbw} and, hence, should not be used in such calculations. 
In addition, when self-consistent calculations are performed, 
it is showed that the magnetic field increase is much slower than an exponential one, not differing by more than one order of magnitude from the surface field. In particular, when calculated for different models and matter compositions, the magnetic field profile inside neutron stars grows  quadratically with baryon chemical potential in the polar direction \cite{Dexheimer:2016yqu}. 
However, determining the magnetic field profile inside neutron stars still needs more refining, as it also depends on the poloidal distribution assumed. Efforts towards the inclusion of combined toroidal \cite{Kiuchi:2008ch,Frieben:2012dz,Ciolfi:2013dta} and poloidal contributions in neutron stars can certainly improve these estimations, even though this is a very challenging task both from  numerical and analytical point of view.

As a final note, we mention that the extent of research topics related to magnetic fields in neutron stars is vast. Although we have used this short proceeding for reviewing only the modeling of equation of state and structure of stars, 
many other topics such as magnetic field impact on crust, thermal evolution and mergers have also interesting open questions  \cite{Negreiros:2018cjk}.



\bibliography{main}

\begin{thebibliography}{49}%
\makeatletter
\providecommand \@ifxundefined [1]{%
 \@ifx{#1\undefined}
}%
\providecommand \@ifnum [1]{%
 \ifnum #1\expandafter \@firstoftwo
 \else \expandafter \@secondoftwo
 \fi
}%
\providecommand \@ifx [1]{%
 \ifx #1\expandafter \@firstoftwo
 \else \expandafter \@secondoftwo
 \fi
}%
\providecommand \natexlab [1]{#1}%
\providecommand \enquote  [1]{``#1''}%
\providecommand \bibnamefont  [1]{#1}%
\providecommand \bibfnamefont [1]{#1}%
\providecommand \citenamefont [1]{#1}%
\providecommand \href@noop [0]{\@secondoftwo}%
\providecommand \href [0]{\begingroup \@sanitize@url \@href}%
\providecommand \@href[1]{\@@startlink{#1}\@@href}%
\providecommand \@@href[1]{\endgroup#1\@@endlink}%
\providecommand \@sanitize@url [0]{\catcode `\\12\catcode `\$12\catcode
  `\&12\catcode `\#12\catcode `\^12\catcode `\_12\catcode `\%12\relax}%
\providecommand \@@startlink[1]{}%
\providecommand \@@endlink[0]{}%
\providecommand \url  [0]{\begingroup\@sanitize@url \@url }%
\providecommand \@url [1]{\endgroup\@href {#1}{\urlprefix }}%
\providecommand \urlprefix  [0]{URL }%
\providecommand \Eprint [0]{\href }%
\providecommand \doibase [0]{http://dx.doi.org/}%
\providecommand \selectlanguage [0]{\@gobble}%
\providecommand \bibinfo  [0]{\@secondoftwo}%
\providecommand \bibfield  [0]{\@secondoftwo}%
\providecommand \translation [1]{[#1]}%
\providecommand \BibitemOpen [0]{}%
\providecommand \bibitemStop [0]{}%
\providecommand \bibitemNoStop [0]{.\EOS\space}%
\providecommand \EOS [0]{\spacefactor3000\relax}%
\providecommand \BibitemShut  [1]{\csname bibitem#1\endcsname}%
\let\auto@bib@innerbib\@empty
\bibitem [{\citenamefont {Duncan}\ and\ \citenamefont
  {Thompson}(1992)}]{Duncan:1992hi}%
  \BibitemOpen
  \bibfield  {author} {\bibinfo {author} {\bibfnamefont {R.~C.}\ \bibnamefont
  {Duncan}}\ and\ \bibinfo {author} {\bibfnamefont {C.}~\bibnamefont
  {Thompson}},\ }\href {\doibase 10.1086/186413} {\bibfield  {journal}
  {\bibinfo  {journal} {Astrophys. J.}\ }\textbf {\bibinfo {volume} {392}},\
  \bibinfo {pages} {L9} (\bibinfo {year} {1992})}\BibitemShut {NoStop}%
\bibitem [{\citenamefont {Kouveliotou}\ \emph {et~al.}(1998)\citenamefont
  {Kouveliotou} \emph {et~al.}}]{Kouveliotou:1998ze}%
  \BibitemOpen
  \bibfield  {author} {\bibinfo {author} {\bibfnamefont {C.}~\bibnamefont
  {Kouveliotou}} \emph {et~al.},\ }\href {\doibase 10.1038/30410} {\bibfield
  {journal} {\bibinfo  {journal} {Nature}\ }\textbf {\bibinfo {volume} {393}},\
  \bibinfo {pages} {235} (\bibinfo {year} {1998})}\BibitemShut {NoStop}%
\bibitem [{\citenamefont {Lewin}\ and\ \citenamefont {van~der
  Klis}(2006)}]{ferro_6}%
  \BibitemOpen
  \bibfield  {author} {\bibinfo {author} {\bibfnamefont {W.}~\bibnamefont
  {Lewin}}\ and\ \bibinfo {author} {\bibfnamefont {M.}~\bibnamefont {van~der
  Klis}},\ }\href@noop {} {\emph {\bibinfo {title} {Compact Stellar X-Ray
  sources}}}\ (\bibinfo  {publisher} {Cambridge University Press},\ \bibinfo
  {year} {2006})\BibitemShut {NoStop}%
\bibitem [{\citenamefont {Livingstone}\ \emph {et~al.}(2007)\citenamefont
  {Livingstone}, \citenamefont {Kaspi}, \citenamefont {Gavriil}, \citenamefont
  {Manchester}, \citenamefont {Gotthelf},\ and\ \citenamefont
  {Kuiper}}]{Livingstone:2007bn}%
  \BibitemOpen
  \bibfield  {author} {\bibinfo {author} {\bibfnamefont {M.~A.}\ \bibnamefont
  {Livingstone}}, \bibinfo {author} {\bibfnamefont {V.~M.}\ \bibnamefont
  {Kaspi}}, \bibinfo {author} {\bibfnamefont {F.~P.}\ \bibnamefont {Gavriil}},
  \bibinfo {author} {\bibfnamefont {R.~N.}\ \bibnamefont {Manchester}},
  \bibinfo {author} {\bibfnamefont {E.~V.}\ \bibnamefont {Gotthelf}}, \ and\
  \bibinfo {author} {\bibfnamefont {L.}~\bibnamefont {Kuiper}},\ }\href
  {\doibase 10.1007/s10509-007-9320-3} {\bibfield  {journal} {\bibinfo
  {journal} {Astrophys. Space Sci.}\ }\textbf {\bibinfo {volume} {308}},\
  \bibinfo {pages} {317} (\bibinfo {year} {2007})},\ \Eprint
  {http://arxiv.org/abs/astro-ph/0702196} {arXiv:astro-ph/0702196 [ASTRO-PH]}
  \BibitemShut {NoStop}%
\bibitem [{\citenamefont {Kaspi}\ and\ \citenamefont
  {Kramer}(2016)}]{Kaspi:2016jkv}%
  \BibitemOpen
  \bibfield  {author} {\bibinfo {author} {\bibfnamefont {V.~M.}\ \bibnamefont
  {Kaspi}}\ and\ \bibinfo {author} {\bibfnamefont {M.}~\bibnamefont {Kramer}}\
  }(\bibinfo {year} {2016})\ \Eprint {http://arxiv.org/abs/1602.07738}
  {arXiv:1602.07738 [astro-ph.HE]} \BibitemShut {NoStop}%
\bibitem [{\citenamefont {Duncan}()}]{duncan_website}%
  \BibitemOpen
  \bibfield  {author} {\bibinfo {author} {\bibfnamefont {R.~C.}\ \bibnamefont
  {Duncan}},\ }\href@noop {} {\bibinfo  {journal}
  {http://solomon.as.utexas.edu/~duncan/magnetar.html}\ }\BibitemShut {NoStop}%
\bibitem [{\citenamefont {Thompson}\ and\ \citenamefont
  {Duncan}(1995)}]{Thompson:1995gw}%
  \BibitemOpen
\bibfield  {journal} {  }\bibfield  {author} {\bibinfo {author} {\bibfnamefont
  {C.}~\bibnamefont {Thompson}}\ and\ \bibinfo {author} {\bibfnamefont {R.~C.}\
  \bibnamefont {Duncan}},\ }\href@noop {} {\bibfield  {journal} {\bibinfo
  {journal} {Mon. Not. Roy. Astron. Soc.}\ }\textbf {\bibinfo {volume} {275}},\
  \bibinfo {pages} {255} (\bibinfo {year} {1995})}\BibitemShut {NoStop}%
\bibitem [{\citenamefont {Thompson}\ and\ \citenamefont
  {Duncan}(1996)}]{Thompson:1996pe}%
  \BibitemOpen
  \bibfield  {author} {\bibinfo {author} {\bibfnamefont {C.}~\bibnamefont
  {Thompson}}\ and\ \bibinfo {author} {\bibfnamefont {R.~C.}\ \bibnamefont
  {Duncan}},\ }\href {\doibase 10.1086/178147} {\bibfield  {journal} {\bibinfo
  {journal} {Astrophys. J.}\ }\textbf {\bibinfo {volume} {473}},\ \bibinfo
  {pages} {322} (\bibinfo {year} {1996})}\BibitemShut {NoStop}%
\bibitem [{\citenamefont {Tatsumi}(2000)}]{Tatsumi:1999ab}%
  \BibitemOpen
  \bibfield  {author} {\bibinfo {author} {\bibfnamefont {T.}~\bibnamefont
  {Tatsumi}},\ }\href {\doibase 10.1016/S0370-2693(00)00927-8} {\bibfield
  {journal} {\bibinfo  {journal} {Phys. Lett.}\ }\textbf {\bibinfo {volume}
  {B489}},\ \bibinfo {pages} {280} (\bibinfo {year} {2000})}\BibitemShut
  {NoStop}%
\bibitem [{\citenamefont {Thompson}\ and\ \citenamefont
  {Duncan}(1993)}]{Thompson:1993hn}%
  \BibitemOpen
  \bibfield  {author} {\bibinfo {author} {\bibfnamefont {C.}~\bibnamefont
  {Thompson}}\ and\ \bibinfo {author} {\bibfnamefont {R.~C.}\ \bibnamefont
  {Duncan}},\ }\href {\doibase 10.1086/172580} {\bibfield  {journal} {\bibinfo
  {journal} {Astrophys. J.}\ }\textbf {\bibinfo {volume} {408}},\ \bibinfo
  {pages} {194} (\bibinfo {year} {1993})}\BibitemShut {NoStop}%
\bibitem [{\citenamefont {Makishima}\ \emph {et~al.}(2014)\citenamefont
  {Makishima}, \citenamefont {Enoto}, \citenamefont {Hiraga}, \citenamefont
  {Nakano}, \citenamefont {Nakazawa}, \citenamefont {Sakurai}, \citenamefont
  {Sasano},\ and\ \citenamefont {Murakami}}]{Makishima:2014dua}%
  \BibitemOpen
  \bibfield  {author} {\bibinfo {author} {\bibfnamefont {K.}~\bibnamefont
  {Makishima}}, \bibinfo {author} {\bibfnamefont {T.}~\bibnamefont {Enoto}},
  \bibinfo {author} {\bibfnamefont {J.~S.}\ \bibnamefont {Hiraga}}, \bibinfo
  {author} {\bibfnamefont {T.}~\bibnamefont {Nakano}}, \bibinfo {author}
  {\bibfnamefont {K.}~\bibnamefont {Nakazawa}}, \bibinfo {author}
  {\bibfnamefont {S.}~\bibnamefont {Sakurai}}, \bibinfo {author} {\bibfnamefont
  {M.}~\bibnamefont {Sasano}}, \ and\ \bibinfo {author} {\bibfnamefont
  {H.}~\bibnamefont {Murakami}},\ }\href {\doibase
  10.1103/PhysRevLett.112.171102} {\bibfield  {journal} {\bibinfo  {journal}
  {Phys. Rev. Lett.}\ }\textbf {\bibinfo {volume} {112}},\ \bibinfo {pages}
  {171102} (\bibinfo {year} {2014})}\BibitemShut {NoStop}%
\bibitem [{\citenamefont {{Lai}}\ and\ \citenamefont
  {{Shapiro}}(1991)}]{Lai1991}%
  \BibitemOpen
  \bibfield  {author} {\bibinfo {author} {\bibfnamefont {D.}~\bibnamefont
  {{Lai}}}\ and\ \bibinfo {author} {\bibfnamefont {S.~L.}\ \bibnamefont
  {{Shapiro}}},\ }\href {\doibase 10.1086/170831} {\bibfield  {journal}
  {\bibinfo  {journal} {Astrophysical Journal}\ }\textbf {\bibinfo {volume}
  {383}},\ \bibinfo {pages} {745} (\bibinfo {year} {1991})}\BibitemShut
  {NoStop}%
\bibitem [{\citenamefont {Cardall}\ \emph {et~al.}(2001)\citenamefont
  {Cardall}, \citenamefont {Prakash},\ and\ \citenamefont
  {Lattimer}}]{Cardall:2000bs}%
  \BibitemOpen
  \bibfield  {author} {\bibinfo {author} {\bibfnamefont {C.~Y.}\ \bibnamefont
  {Cardall}}, \bibinfo {author} {\bibfnamefont {M.}~\bibnamefont {Prakash}}, \
  and\ \bibinfo {author} {\bibfnamefont {J.~M.}\ \bibnamefont {Lattimer}},\
  }\href {\doibase 10.1086/321370} {\bibfield  {journal} {\bibinfo  {journal}
  {Astrophys. J.}\ }\textbf {\bibinfo {volume} {554}},\ \bibinfo {pages} {322}
  (\bibinfo {year} {2001})}\BibitemShut {NoStop}%
\bibitem [{\citenamefont {Ferrer}\ \emph {et~al.}(2010)\citenamefont {Ferrer},
  \citenamefont {de~la Incera}, \citenamefont {Keith}, \citenamefont
  {Portillo},\ and\ \citenamefont {Springsteen}}]{Ferrer:2010wz}%
  \BibitemOpen
  \bibfield  {author} {\bibinfo {author} {\bibfnamefont {E.~J.}\ \bibnamefont
  {Ferrer}}, \bibinfo {author} {\bibfnamefont {V.}~\bibnamefont {de~la
  Incera}}, \bibinfo {author} {\bibfnamefont {J.~P.}\ \bibnamefont {Keith}},
  \bibinfo {author} {\bibfnamefont {I.}~\bibnamefont {Portillo}}, \ and\
  \bibinfo {author} {\bibfnamefont {P.~L.}\ \bibnamefont {Springsteen}},\
  }\href {\doibase 10.1103/PhysRevC.82.065802} {\bibfield  {journal} {\bibinfo
  {journal} {Phys. Rev.}\ }\textbf {\bibinfo {volume} {C82}},\ \bibinfo {pages}
  {065802} (\bibinfo {year} {2010})}\BibitemShut {NoStop}%
\bibitem [{\citenamefont {Negreiros}\ \emph {et~al.}(2018)\citenamefont
  {Negreiros}, \citenamefont {Bernal}, \citenamefont {Dexheimer},\ and\
  \citenamefont {Troconis}}]{Negreiros:2018cjk}%
  \BibitemOpen
  \bibfield  {author} {\bibinfo {author} {\bibfnamefont {R.}~\bibnamefont
  {Negreiros}}, \bibinfo {author} {\bibfnamefont {C.}~\bibnamefont {Bernal}},
  \bibinfo {author} {\bibfnamefont {V.}~\bibnamefont {Dexheimer}}, \ and\
  \bibinfo {author} {\bibfnamefont {O.}~\bibnamefont {Troconis}},\ }\href
  {\doibase 10.3390/universe4030043} {\bibfield  {journal} {\bibinfo  {journal}
  {Universe}\ }\textbf {\bibinfo {volume} {4}},\ \bibinfo {pages} {43}
  (\bibinfo {year} {2018})}\BibitemShut {NoStop}%
\bibitem [{\citenamefont {Chiu}\ \emph {et~al.}(1968)\citenamefont {Chiu},
  \citenamefont {Canuto},\ and\ \citenamefont {Fassio-Canuto}}]{Chiu:1968zz}%
  \BibitemOpen
  \bibfield  {author} {\bibinfo {author} {\bibfnamefont {H.-Y.}\ \bibnamefont
  {Chiu}}, \bibinfo {author} {\bibfnamefont {V.}~\bibnamefont {Canuto}}, \ and\
  \bibinfo {author} {\bibfnamefont {L.}~\bibnamefont {Fassio-Canuto}},\ }\href
  {\doibase 10.1103/PhysRev.176.1438} {\bibfield  {journal} {\bibinfo
  {journal} {Phys. Rev.}\ }\textbf {\bibinfo {volume} {176}},\ \bibinfo {pages}
  {1438} (\bibinfo {year} {1968})}\BibitemShut {NoStop}%
\bibitem [{\citenamefont {Canuto}\ and\ \citenamefont
  {Chiu}(1968{\natexlab{a}})}]{Canuto:1969ct}%
  \BibitemOpen
  \bibfield  {author} {\bibinfo {author} {\bibfnamefont {V.}~\bibnamefont
  {Canuto}}\ and\ \bibinfo {author} {\bibfnamefont {H.~Y.}\ \bibnamefont
  {Chiu}},\ }\href {\doibase 10.1103/PhysRev.173.1210} {\bibfield  {journal}
  {\bibinfo  {journal} {Phys. Rev.}\ }\textbf {\bibinfo {volume} {173}},\
  \bibinfo {pages} {1210} (\bibinfo {year} {1968}{\natexlab{a}})}\BibitemShut
  {NoStop}%
\bibitem [{\citenamefont {Canuto}\ and\ \citenamefont
  {Chiu}(1968{\natexlab{b}})}]{Canuto:1969cs}%
  \BibitemOpen
  \bibfield  {author} {\bibinfo {author} {\bibfnamefont {V.}~\bibnamefont
  {Canuto}}\ and\ \bibinfo {author} {\bibfnamefont {H.~Y.}\ \bibnamefont
  {Chiu}},\ }\href {\doibase 10.1103/PhysRev.173.1220} {\bibfield  {journal}
  {\bibinfo  {journal} {Phys. Rev.}\ }\textbf {\bibinfo {volume} {173}},\
  \bibinfo {pages} {1220} (\bibinfo {year} {1968}{\natexlab{b}})}\BibitemShut
  {NoStop}%
\bibitem [{\citenamefont {Canuto}\ and\ \citenamefont
  {Chiu}(1968{\natexlab{c}})}]{Canuto:1969cn}%
  \BibitemOpen
  \bibfield  {author} {\bibinfo {author} {\bibfnamefont {V.}~\bibnamefont
  {Canuto}}\ and\ \bibinfo {author} {\bibfnamefont {H.~Y.}\ \bibnamefont
  {Chiu}},\ }\href {\doibase 10.1103/PhysRev.173.1229} {\bibfield  {journal}
  {\bibinfo  {journal} {Phys. Rev.}\ }\textbf {\bibinfo {volume} {173}},\
  \bibinfo {pages} {1229} (\bibinfo {year} {1968}{\natexlab{c}})}\BibitemShut
  {NoStop}%
\bibitem [{\citenamefont {Chakrabarty}\ \emph {et~al.}(1997)\citenamefont
  {Chakrabarty}, \citenamefont {Bandyopadhyay},\ and\ \citenamefont
  {Pal}}]{Chakrabarty:1997ef}%
  \BibitemOpen
  \bibfield  {author} {\bibinfo {author} {\bibfnamefont {S.}~\bibnamefont
  {Chakrabarty}}, \bibinfo {author} {\bibfnamefont {D.}~\bibnamefont
  {Bandyopadhyay}}, \ and\ \bibinfo {author} {\bibfnamefont {S.}~\bibnamefont
  {Pal}},\ }\href {\doibase 10.1103/PhysRevLett.78.2898} {\bibfield  {journal}
  {\bibinfo  {journal} {Phys. Rev. Lett.}\ }\textbf {\bibinfo {volume} {78}},\
  \bibinfo {pages} {2898} (\bibinfo {year} {1997})}\BibitemShut {NoStop}%
\bibitem [{\citenamefont {Broderick}\ \emph {et~al.}(2000)\citenamefont
  {Broderick}, \citenamefont {Prakash},\ and\ \citenamefont
  {Lattimer}}]{Broderick:2000pe}%
  \BibitemOpen
  \bibfield  {author} {\bibinfo {author} {\bibfnamefont {A.~E.}\ \bibnamefont
  {Broderick}}, \bibinfo {author} {\bibfnamefont {M.}~\bibnamefont {Prakash}},
  \ and\ \bibinfo {author} {\bibfnamefont {J.~M.}\ \bibnamefont {Lattimer}},\
  }\href@noop {} {\bibfield  {journal} {\bibinfo  {journal} {Astrophys. J.}\
  }\textbf {\bibinfo {volume} {537}},\ \bibinfo {pages} {351} (\bibinfo {year}
  {2000})}\BibitemShut {NoStop}%
\bibitem [{\citenamefont {Strickland}\ \emph {et~al.}(2012)\citenamefont
  {Strickland}, \citenamefont {Dexheimer},\ and\ \citenamefont
  {Menezes}}]{Strickland:2012vu}%
  \BibitemOpen
  \bibfield  {author} {\bibinfo {author} {\bibfnamefont {M.}~\bibnamefont
  {Strickland}}, \bibinfo {author} {\bibfnamefont {V.}~\bibnamefont
  {Dexheimer}}, \ and\ \bibinfo {author} {\bibfnamefont {D.~P.}\ \bibnamefont
  {Menezes}},\ }\href {\doibase 10.1103/PhysRevD.86.125032} {\bibfield
  {journal} {\bibinfo  {journal} {Phys. Rev.}\ }\textbf {\bibinfo {volume}
  {D86}},\ \bibinfo {pages} {125032} (\bibinfo {year} {2012})}\BibitemShut
  {NoStop}%
\bibitem [{\citenamefont {Broderick}\ \emph {et~al.}(2002)\citenamefont
  {Broderick}, \citenamefont {Prakash},\ and\ \citenamefont
  {Lattimer}}]{Broderick:2001qw}%
  \BibitemOpen
  \bibfield  {author} {\bibinfo {author} {\bibfnamefont {A.~E.}\ \bibnamefont
  {Broderick}}, \bibinfo {author} {\bibfnamefont {M.}~\bibnamefont {Prakash}},
  \ and\ \bibinfo {author} {\bibfnamefont {J.~M.}\ \bibnamefont {Lattimer}},\
  }\href@noop {} {\bibfield  {journal} {\bibinfo  {journal} {Phys. Lett.}\
  }\textbf {\bibinfo {volume} {B531}},\ \bibinfo {pages} {167} (\bibinfo {year}
  {2002})}\BibitemShut {NoStop}%
\bibitem [{\citenamefont {Perez~Martinez}\ \emph {et~al.}(2008)\citenamefont
  {Perez~Martinez}, \citenamefont {Perez~Rojas},\ and\ \citenamefont
  {Mosquera~Cuesta}}]{PerezMartinez:2007kw}%
  \BibitemOpen
  \bibfield  {author} {\bibinfo {author} {\bibfnamefont {A.}~\bibnamefont
  {Perez~Martinez}}, \bibinfo {author} {\bibfnamefont {H.}~\bibnamefont
  {Perez~Rojas}}, \ and\ \bibinfo {author} {\bibfnamefont {H.}~\bibnamefont
  {Mosquera~Cuesta}},\ }\href {\doibase 10.1142/S0218271808013741} {\bibfield
  {journal} {\bibinfo  {journal} {Int. J. Mod. Phys.}\ }\textbf {\bibinfo
  {volume} {D17}},\ \bibinfo {pages} {2107} (\bibinfo {year}
  {2008})}\BibitemShut {NoStop}%
\bibitem [{\citenamefont {Dexheimer}\ \emph {et~al.}(2014)\citenamefont
  {Dexheimer}, \citenamefont {Menezes},\ and\ \citenamefont
  {Strickland}}]{Dexheimer:2012mk}%
  \BibitemOpen
  \bibfield  {author} {\bibinfo {author} {\bibfnamefont {V.}~\bibnamefont
  {Dexheimer}}, \bibinfo {author} {\bibfnamefont {D.~P.}\ \bibnamefont
  {Menezes}}, \ and\ \bibinfo {author} {\bibfnamefont {M.}~\bibnamefont
  {Strickland}},\ }\href {\doibase 10.1088/0954-3899/41/1/015203} {\bibfield
  {journal} {\bibinfo  {journal} {J. Phys.}\ }\textbf {\bibinfo {volume}
  {G41}},\ \bibinfo {pages} {015203} (\bibinfo {year} {2014})}\BibitemShut
  {NoStop}%
\bibitem [{\citenamefont {Dexheimer}\ \emph {et~al.}(2013)\citenamefont
  {Dexheimer}, \citenamefont {Negreiros}, \citenamefont {Schramm},\ and\
  \citenamefont {Hempel}}]{Dexheimer:2012qk}%
  \BibitemOpen
  \bibfield  {author} {\bibinfo {author} {\bibfnamefont {V.}~\bibnamefont
  {Dexheimer}}, \bibinfo {author} {\bibfnamefont {R.}~\bibnamefont
  {Negreiros}}, \bibinfo {author} {\bibfnamefont {S.}~\bibnamefont {Schramm}},
  \ and\ \bibinfo {author} {\bibfnamefont {M.}~\bibnamefont {Hempel}},\
  }\href@noop {} {\bibfield  {journal} {\bibinfo  {journal} {AIP Conf. Proc.}\
  }\textbf {\bibinfo {volume} {1520}},\ \bibinfo {pages} {264} (\bibinfo {year}
  {2013})}\BibitemShut {NoStop}%
\bibitem [{\citenamefont {Ferrer}\ \emph {et~al.}(2015)\citenamefont {Ferrer},
  \citenamefont {de~la Incera}, \citenamefont {Manreza~Paret}, \citenamefont
  {Pérez~Martínez},\ and\ \citenamefont {Sanchez}}]{Ferrer:2015wca}%
  \BibitemOpen
  \bibfield  {author} {\bibinfo {author} {\bibfnamefont {E.~J.}\ \bibnamefont
  {Ferrer}}, \bibinfo {author} {\bibfnamefont {V.}~\bibnamefont {de~la
  Incera}}, \bibinfo {author} {\bibfnamefont {D.}~\bibnamefont
  {Manreza~Paret}}, \bibinfo {author} {\bibfnamefont {A.}~\bibnamefont
  {Pérez~Martínez}}, \ and\ \bibinfo {author} {\bibfnamefont
  {A.}~\bibnamefont {Sanchez}},\ }\href {\doibase 10.1103/PhysRevD.91.085041}
  {\bibfield  {journal} {\bibinfo  {journal} {Phys. Rev.}\ }\textbf {\bibinfo
  {volume} {D91}},\ \bibinfo {pages} {085041} (\bibinfo {year}
  {2015})}\BibitemShut {NoStop}%
\bibitem [{\citenamefont {Rabhi}\ \emph {et~al.}(2009)\citenamefont {Rabhi},
  \citenamefont {Pais}, \citenamefont {Panda},\ and\ \citenamefont
  {Providencia}}]{Rabhi:2009ih}%
  \BibitemOpen
  \bibfield  {author} {\bibinfo {author} {\bibfnamefont {A.}~\bibnamefont
  {Rabhi}}, \bibinfo {author} {\bibfnamefont {H.}~\bibnamefont {Pais}},
  \bibinfo {author} {\bibfnamefont {P.~K.}\ \bibnamefont {Panda}}, \ and\
  \bibinfo {author} {\bibfnamefont {C.}~\bibnamefont {Providencia}},\ }\href
  {\doibase 10.1088/0954-3899/36/11/115204} {\bibfield  {journal} {\bibinfo
  {journal} {J. Phys.}\ }\textbf {\bibinfo {volume} {G36}},\ \bibinfo {pages}
  {115204} (\bibinfo {year} {2009})}\BibitemShut {NoStop}%
\bibitem [{\citenamefont {Paulucci}\ \emph {et~al.}(2011)\citenamefont
  {Paulucci}, \citenamefont {Ferrer}, \citenamefont {de~la Incera},\ and\
  \citenamefont {Horvath}}]{Paulucci:2010uj}%
  \BibitemOpen
  \bibfield  {author} {\bibinfo {author} {\bibfnamefont {L.}~\bibnamefont
  {Paulucci}}, \bibinfo {author} {\bibfnamefont {E.~J.}\ \bibnamefont
  {Ferrer}}, \bibinfo {author} {\bibfnamefont {V.}~\bibnamefont {de~la
  Incera}}, \ and\ \bibinfo {author} {\bibfnamefont {J.~E.}\ \bibnamefont
  {Horvath}},\ }\href {\doibase 10.1103/PhysRevD.83.043009} {\bibfield
  {journal} {\bibinfo  {journal} {Phys. Rev.}\ }\textbf {\bibinfo {volume}
  {D83}},\ \bibinfo {pages} {043009} (\bibinfo {year} {2011})}\BibitemShut
  {NoStop}%
\bibitem [{\citenamefont {Sinha}\ \emph {et~al.}(2013)\citenamefont {Sinha},
  \citenamefont {Mukhopadhyay},\ and\ \citenamefont
  {Sedrakian}}]{Sinha:2010fm}%
  \BibitemOpen
  \bibfield  {author} {\bibinfo {author} {\bibfnamefont {M.}~\bibnamefont
  {Sinha}}, \bibinfo {author} {\bibfnamefont {B.}~\bibnamefont {Mukhopadhyay}},
  \ and\ \bibinfo {author} {\bibfnamefont {A.}~\bibnamefont {Sedrakian}},\
  }\href {\doibase 10.1016/j.nuclphysa.2012.12.076} {\bibfield  {journal}
  {\bibinfo  {journal} {Nucl. Phys.}\ }\textbf {\bibinfo {volume} {A898}},\
  \bibinfo {pages} {43} (\bibinfo {year} {2013})}\BibitemShut {NoStop}%
\bibitem [{\citenamefont {Orsaria}\ \emph {et~al.}(2011)\citenamefont
  {Orsaria}, \citenamefont {Ranea-Sandoval},\ and\ \citenamefont
  {Vucetich}}]{Orsaria:2010xx}%
  \BibitemOpen
  \bibfield  {author} {\bibinfo {author} {\bibfnamefont {M.}~\bibnamefont
  {Orsaria}}, \bibinfo {author} {\bibfnamefont {I.~F.}\ \bibnamefont
  {Ranea-Sandoval}}, \ and\ \bibinfo {author} {\bibfnamefont {H.}~\bibnamefont
  {Vucetich}},\ }\href {\doibase 10.1088/0004-637X/734/1/41} {\bibfield
  {journal} {\bibinfo  {journal} {Astrophys. J.}\ }\textbf {\bibinfo {volume}
  {734}},\ \bibinfo {pages} {41} (\bibinfo {year} {2011})}\BibitemShut
  {NoStop}%
\bibitem [{\citenamefont {Dexheimer}\ \emph {et~al.}(2012)\citenamefont
  {Dexheimer}, \citenamefont {Negreiros},\ and\ \citenamefont
  {Schramm}}]{Dexheimer:2011pz}%
  \BibitemOpen
  \bibfield  {author} {\bibinfo {author} {\bibfnamefont {V.}~\bibnamefont
  {Dexheimer}}, \bibinfo {author} {\bibfnamefont {R.}~\bibnamefont
  {Negreiros}}, \ and\ \bibinfo {author} {\bibfnamefont {S.}~\bibnamefont
  {Schramm}},\ }\href {\doibase 10.1140/epja/i2012-12189-y} {\bibfield
  {journal} {\bibinfo  {journal} {Eur. Phys. J.}\ }\textbf {\bibinfo {volume}
  {A48}},\ \bibinfo {pages} {189} (\bibinfo {year} {2012})}\BibitemShut
  {NoStop}%
\bibitem [{\citenamefont {Lopes}\ and\ \citenamefont
  {Menezes}(2012)}]{Lopes:2012nf}%
  \BibitemOpen
  \bibfield  {author} {\bibinfo {author} {\bibfnamefont {L.~L.}\ \bibnamefont
  {Lopes}}\ and\ \bibinfo {author} {\bibfnamefont {D.~P.}\ \bibnamefont
  {Menezes}},\ }\href {\doibase 10.1007/s13538-012-0093-y} {\bibfield
  {journal} {\bibinfo  {journal} {Brazilian Journal of Physics 42}\ ,\ \bibinfo
  {pages} {428}} (\bibinfo {year} {2012})}\BibitemShut {NoStop}%
\bibitem [{\citenamefont {Casali}\ \emph {et~al.}(2014)\citenamefont {Casali},
  \citenamefont {Castro},\ and\ \citenamefont {Menezes}}]{Casali:2013jka}%
  \BibitemOpen
  \bibfield  {author} {\bibinfo {author} {\bibfnamefont {R.~H.}\ \bibnamefont
  {Casali}}, \bibinfo {author} {\bibfnamefont {L.~B.}\ \bibnamefont {Castro}},
  \ and\ \bibinfo {author} {\bibfnamefont {D.~P.}\ \bibnamefont {Menezes}},\
  }\href@noop {} {\bibfield  {journal} {\bibinfo  {journal} {Phys. Rev.}\
  }\textbf {\bibinfo {volume} {C89}},\ \bibinfo {pages} {015805} (\bibinfo
  {year} {2014})}\BibitemShut {NoStop}%
\bibitem [{\citenamefont {Denke}\ and\ \citenamefont
  {Pinto}(2013)}]{Denke:2013gha}%
  \BibitemOpen
  \bibfield  {author} {\bibinfo {author} {\bibfnamefont {R.~Z.}\ \bibnamefont
  {Denke}}\ and\ \bibinfo {author} {\bibfnamefont {M.~B.}\ \bibnamefont
  {Pinto}},\ }\href {\doibase 10.1103/PhysRevD.88.056008} {\bibfield  {journal}
  {\bibinfo  {journal} {Phys. Rev.}\ }\textbf {\bibinfo {volume} {D88}},\
  \bibinfo {pages} {056008} (\bibinfo {year} {2013})}\BibitemShut {NoStop}%
\bibitem [{\citenamefont {Gomes}\ \emph {et~al.}(2014)\citenamefont {Gomes},
  \citenamefont {Dexheimer},\ and\ \citenamefont
  {Vasconcellos}}]{Gomes:2014dka}%
  \BibitemOpen
  \bibfield  {author} {\bibinfo {author} {\bibfnamefont {R.~O.}\ \bibnamefont
  {Gomes}}, \bibinfo {author} {\bibfnamefont {V.}~\bibnamefont {Dexheimer}}, \
  and\ \bibinfo {author} {\bibfnamefont {C.~A.~Z.}\ \bibnamefont
  {Vasconcellos}},\ }\bibfield  {booktitle} {\emph {\bibinfo {booktitle}
  {{Astron.Nachr. 335 (2014) 666}}},\ }\href {\doibase 10.1002/asna.201412090}
  {\bibfield  {journal} {\bibinfo  {journal} {Astron. Nachr.}\ }\textbf
  {\bibinfo {volume} {335}},\ \bibinfo {pages} {666} (\bibinfo {year}
  {2014})}\BibitemShut {NoStop}%
\bibitem [{\citenamefont {Schramm}\ \emph {et~al.}(2015)\citenamefont
  {Schramm}, \citenamefont {Bhattacharyya}, \citenamefont {Dexheimer},\ and\
  \citenamefont {Mallick}}]{Schramm:2015lga}%
  \BibitemOpen
  \bibfield  {author} {\bibinfo {author} {\bibfnamefont {S.}~\bibnamefont
  {Schramm}}, \bibinfo {author} {\bibfnamefont {A.}~\bibnamefont
  {Bhattacharyya}}, \bibinfo {author} {\bibfnamefont {V.}~\bibnamefont
  {Dexheimer}}, \ and\ \bibinfo {author} {\bibfnamefont {R.}~\bibnamefont
  {Mallick}},\ }in\ \href@noop {} {\emph {\bibinfo {booktitle} {{CSQCD IV,
  Prerow, Germany, }}}}\ (\bibinfo {year} {2015})\ \Eprint
  {http://arxiv.org/abs/1504.00451} {arXiv:1504.00451 [astro-ph.SR]}
  \BibitemShut {NoStop}%
\bibitem [{\citenamefont {Gomes}\ \emph {et~al.}(2013)\citenamefont {Gomes},
  \citenamefont {Dexheimer},\ and\ \citenamefont
  {Vasconcellos}}]{Gomes:2013sra}%
  \BibitemOpen
  \bibfield  {author} {\bibinfo {author} {\bibfnamefont {R.~O.}\ \bibnamefont
  {Gomes}}, \bibinfo {author} {\bibfnamefont {V.}~\bibnamefont {Dexheimer}}, \
  and\ \bibinfo {author} {\bibfnamefont {C.~A.~Z.}\ \bibnamefont
  {Vasconcellos}},\ }in\ \href
  {https://inspirehep.net/record/1244976/files/arXiv:1307.7450.pdf} {\emph
  {\bibinfo {booktitle} {{Compact Stars in the QCD Phase Diagram III (CSQCD
  III) Guarujá, SP, Brazil, December 12-15, 2012}}}}\ (\bibinfo {year}
  {2013})\ \Eprint {http://arxiv.org/abs/1307.7450} {arXiv:1307.7450 [nucl-th]}
  \BibitemShut {NoStop}%
\bibitem [{\citenamefont {Gomes}\ \emph {et~al.}(2017)\citenamefont {Gomes},
  \citenamefont {Franzon}, \citenamefont {Dexheimer},\ and\ \citenamefont
  {Schramm}}]{Gomes:2017zkc}%
  \BibitemOpen
  \bibfield  {author} {\bibinfo {author} {\bibfnamefont {R.~O.}\ \bibnamefont
  {Gomes}}, \bibinfo {author} {\bibfnamefont {B.}~\bibnamefont {Franzon}},
  \bibinfo {author} {\bibfnamefont {V.}~\bibnamefont {Dexheimer}}, \ and\
  \bibinfo {author} {\bibfnamefont {S.}~\bibnamefont {Schramm}},\ }\href
  {\doibase 10.3847/1538-4357/aa8b68} {\bibfield  {journal} {\bibinfo
  {journal} {Astrophys. J.}\ }\textbf {\bibinfo {volume} {850}},\ \bibinfo
  {pages} {20} (\bibinfo {year} {2017})}\BibitemShut {NoStop}%
\bibitem [{\citenamefont {Franzon}\ \emph {et~al.}(2015)\citenamefont
  {Franzon}, \citenamefont {Dexheimer},\ and\ \citenamefont
  {Schramm}}]{Franzon:2015sya}%
  \BibitemOpen
  \bibfield  {author} {\bibinfo {author} {\bibfnamefont {B.}~\bibnamefont
  {Franzon}}, \bibinfo {author} {\bibfnamefont {V.}~\bibnamefont {Dexheimer}},
  \ and\ \bibinfo {author} {\bibfnamefont {S.}~\bibnamefont {Schramm}},\ }\href
  {\doibase 10.1093/mnras/stv2606} {\bibfield  {journal} {\bibinfo  {journal}
  {Mon. Not. Roy. Astron. Soc.}\ }\textbf {\bibinfo {volume} {456}},\ \bibinfo
  {pages} {2937} (\bibinfo {year} {2015})}\BibitemShut {NoStop}%
\bibitem [{\citenamefont {Franzon}\ \emph {et~al.}(2016)\citenamefont
  {Franzon}, \citenamefont {Gomes},\ and\ \citenamefont
  {Schramm}}]{Franzon:2016urz}%
  \BibitemOpen
  \bibfield  {author} {\bibinfo {author} {\bibfnamefont {B.}~\bibnamefont
  {Franzon}}, \bibinfo {author} {\bibfnamefont {R.~O.}\ \bibnamefont {Gomes}},
  \ and\ \bibinfo {author} {\bibfnamefont {S.}~\bibnamefont {Schramm}},\ }\href
  {\doibase 10.1093/mnras/stw1967} {\bibfield  {journal} {\bibinfo  {journal}
  {Mon. Not. Roy. Astron. Soc.}\ }\textbf {\bibinfo {volume} {463}},\ \bibinfo
  {pages} {571} (\bibinfo {year} {2016})}\BibitemShut {NoStop}%
\bibitem [{\citenamefont {Bonazzola}\ \emph {et~al.}(1993)\citenamefont
  {Bonazzola}, \citenamefont {Gourgoulhon}, \citenamefont {Salgado},\ and\
  \citenamefont {Marck}}]{Bonazzola:1993zz}%
  \BibitemOpen
  \bibfield  {author} {\bibinfo {author} {\bibfnamefont {S.}~\bibnamefont
  {Bonazzola}}, \bibinfo {author} {\bibfnamefont {E.}~\bibnamefont
  {Gourgoulhon}}, \bibinfo {author} {\bibfnamefont {M.}~\bibnamefont
  {Salgado}}, \ and\ \bibinfo {author} {\bibfnamefont {J.~A.}\ \bibnamefont
  {Marck}},\ }\href@noop {} {\bibfield  {journal} {\bibinfo  {journal} {Astron.
  Astrophys.}\ }\textbf {\bibinfo {volume} {278}},\ \bibinfo {pages} {421}
  (\bibinfo {year} {1993})}\BibitemShut {NoStop}%
\bibitem [{\citenamefont {Bocquet}\ \emph {et~al.}(1995)\citenamefont
  {Bocquet}, \citenamefont {Bonazzola}, \citenamefont {Gourgoulhon},\ and\
  \citenamefont {Novak}}]{Bocquet:1995je}%
  \BibitemOpen
  \bibfield  {author} {\bibinfo {author} {\bibfnamefont {M.}~\bibnamefont
  {Bocquet}}, \bibinfo {author} {\bibfnamefont {S.}~\bibnamefont {Bonazzola}},
  \bibinfo {author} {\bibfnamefont {E.}~\bibnamefont {Gourgoulhon}}, \ and\
  \bibinfo {author} {\bibfnamefont {J.}~\bibnamefont {Novak}},\ }\href@noop {}
  {\bibfield  {journal} {\bibinfo  {journal} {Astron. Astrophys.}\ }\textbf
  {\bibinfo {volume} {301}},\ \bibinfo {pages} {757} (\bibinfo {year}
  {1995})}\BibitemShut {NoStop}%
\bibitem [{\citenamefont {Chatterjee}\ \emph {et~al.}(2015)\citenamefont
  {Chatterjee}, \citenamefont {Elghozi}, \citenamefont {Novak},\ and\
  \citenamefont {Oertel}}]{Chatterjee:2014qsa}%
  \BibitemOpen
  \bibfield  {author} {\bibinfo {author} {\bibfnamefont {D.}~\bibnamefont
  {Chatterjee}}, \bibinfo {author} {\bibfnamefont {T.}~\bibnamefont {Elghozi}},
  \bibinfo {author} {\bibfnamefont {J.}~\bibnamefont {Novak}}, \ and\ \bibinfo
  {author} {\bibfnamefont {M.}~\bibnamefont {Oertel}},\ }\href {\doibase
  10.1093/mnras/stu2706} {\bibfield  {journal} {\bibinfo  {journal} {Mon. Not.
  Roy. Astron. Soc.}\ }\textbf {\bibinfo {volume} {447}},\ \bibinfo {pages}
  {3785} (\bibinfo {year} {2015})}\BibitemShut {NoStop}%
\bibitem [{\citenamefont {Dexheimer}\ \emph {et~al.}(2017)\citenamefont
  {Dexheimer}, \citenamefont {Franzon}, \citenamefont {Gomes}, \citenamefont
  {Farias}, \citenamefont {Avancini},\ and\ \citenamefont
  {Schramm}}]{Dexheimer:2016yqu}%
  \BibitemOpen
  \bibfield  {author} {\bibinfo {author} {\bibfnamefont {V.}~\bibnamefont
  {Dexheimer}}, \bibinfo {author} {\bibfnamefont {B.}~\bibnamefont {Franzon}},
  \bibinfo {author} {\bibfnamefont {R.~O.}\ \bibnamefont {Gomes}}, \bibinfo
  {author} {\bibfnamefont {R.~L.~S.}\ \bibnamefont {Farias}}, \bibinfo {author}
  {\bibfnamefont {S.~S.}\ \bibnamefont {Avancini}}, \ and\ \bibinfo {author}
  {\bibfnamefont {S.}~\bibnamefont {Schramm}},\ }\href {\doibase
  10.1016/j.physletb.2017.09.008} {\bibfield  {journal} {\bibinfo  {journal}
  {Phys. Lett.}\ }\textbf {\bibinfo {volume} {B773}},\ \bibinfo {pages} {487}
  (\bibinfo {year} {2017})}\BibitemShut {NoStop}%
\bibitem [{\citenamefont {Alloy}\ and\ \citenamefont
  {Menezes}(2017)}]{Menezes:2016wbw}%
  \BibitemOpen
  \bibfield  {author} {\bibinfo {author} {\bibfnamefont {M.~D.}\ \bibnamefont
  {Alloy}}\ and\ \bibinfo {author} {\bibfnamefont {D.~P.}\ \bibnamefont
  {Menezes}},\ }\href {\doibase 10.1142/S201019451760031X} {\bibfield
  {journal} {\bibinfo  {journal} {Int. J. Mod. Phys. Conf. Ser.}\ }\textbf
  {\bibinfo {volume} {45}},\ \bibinfo {pages} {1760031} (\bibinfo {year}
  {2017})}\BibitemShut {NoStop}%
\bibitem [{\citenamefont {Kiuchi}\ and\ \citenamefont
  {Yoshida}(2008)}]{Kiuchi:2008ch}%
  \BibitemOpen
  \bibfield  {author} {\bibinfo {author} {\bibfnamefont {K.}~\bibnamefont
  {Kiuchi}}\ and\ \bibinfo {author} {\bibfnamefont {S.}~\bibnamefont
  {Yoshida}},\ }\href {\doibase 10.1103/PhysRevD.78.044045} {\bibfield
  {journal} {\bibinfo  {journal} {Phys. Rev.}\ }\textbf {\bibinfo {volume}
  {D78}},\ \bibinfo {pages} {044045} (\bibinfo {year} {2008})},\ \Eprint
  {http://arxiv.org/abs/0802.2983} {arXiv:0802.2983 [astro-ph]} \BibitemShut
  {NoStop}%
\bibitem [{\citenamefont {Frieben}\ and\ \citenamefont
  {Rezzolla}(2012)}]{Frieben:2012dz}%
  \BibitemOpen
  \bibfield  {author} {\bibinfo {author} {\bibfnamefont {J.}~\bibnamefont
  {Frieben}}\ and\ \bibinfo {author} {\bibfnamefont {L.}~\bibnamefont
  {Rezzolla}},\ }\href {\doibase 10.1111/j.1365-2966.2012.22027.x} {\bibfield
  {journal} {\bibinfo  {journal} {Mon. Not. Roy. Astron. Soc.}\ }\textbf
  {\bibinfo {volume} {427}},\ \bibinfo {pages} {3406} (\bibinfo {year}
  {2012})}\BibitemShut {NoStop}%
\bibitem [{\citenamefont {Ciolfi}\ and\ \citenamefont
  {Rezzolla}(2013)}]{Ciolfi:2013dta}%
  \BibitemOpen
  \bibfield  {author} {\bibinfo {author} {\bibfnamefont {R.}~\bibnamefont
  {Ciolfi}}\ and\ \bibinfo {author} {\bibfnamefont {L.}~\bibnamefont
  {Rezzolla}},\ }\href {\doibase 10.1093/mnrasl/slt092} {\bibfield  {journal}
  {\bibinfo  {journal} {Mon. Not. Roy. Astron. Soc.}\ }\textbf {\bibinfo
  {volume} {435}},\ \bibinfo {pages} {L43} (\bibinfo {year} {2013})},\ \Eprint
  {http://arxiv.org/abs/1306.2803} {arXiv:1306.2803 [astro-ph.SR]} \BibitemShut
  {NoStop}%
\end{thebibliography}%

\end{document}